\documentclass[useAMS,usenatbib]{mn2e}
\usepackage{graphicx,times}
\usepackage{natbib}
\usepackage{amssymb}

\title[Simulations of the merging galaxy cluster Abell~3376]{Simulations of the merging galaxy cluster Abell~3376}

\author[R. E. G. Machado \& G. B. Lima Neto]{Rubens E. G. Machado$^{1}$\thanks{E-mail: rgmachado@astro.iag.usp.br} and Gast\~ao B. Lima Neto$^{1}$\\ $^{1}$Instituto de Astronomia, Geof\'isica e Ci\^encias Atmosf\'ericas, Universidade de S\~ao Paulo, R. do Mat\~ao 1226, 05508-090 S\~ao Paulo, Brazil}

\begin{document}  

\date{Accepted 2013 January 18. Received 2013 January 10; in original form 2012 October 25}

\pagerange{\pageref{firstpage}--\pageref{lastpage}} \pubyear{2012}

\maketitle

\label{firstpage}

\begin{abstract}
Observed galaxy clusters often exhibit X-ray morphologies suggestive of recent interaction with an infalling subcluster. Abell~3376 is a nearby ($z=0.046$) massive galaxy cluster whose bullet-shaped X-ray emission indicates that it may have undergone a recent collision. It displays a pair of Mpc-scale radio relics and its brightest cluster galaxy is located $970 h_{70}^{-1}\,$kpc away from the peak of X-ray emission, where the second brightest galaxy lies.
We attempt to recover the dynamical history of Abell~3376.
We perform a set of $N$-body adiabatic hydrodynamical simulations using the SPH code Gadget-2. These simulations of binary cluster collisions are aimed at exploring the parameter space of possible initial configurations. By attempting to match X-ray morphology, temperature, virial mass and X-ray luminosity, we set approximate constraints on some merger parameters.
Our best models suggest a collision of clusters with mass ratio in the range 1/6--1/8, and having a subcluster with central gas density four times higher than that of the major cluster. Models with small impact parameter ($b<150$~kpc), if any, are preferred. We estimate that Abell~3376 is observed approximately 0.5~Gyr after core passage, and that the collision axis is inclined by $i\approx40^{\circ}$ with respect to the plane of the sky. The infalling subcluster drives a supersonic shock wave that propagates at almost 2600~km/s, implying a Mach number as high as $\mathcal{M}\sim4$; but we show how it would have been underestimated as $\mathcal{M}\sim3$ due to projection effects.
\end{abstract}



\section{Introduction}
\label{sec:intro}

Mergers are the processes through which galaxy clusters assemble in the hierarchical scenario of structure formation. Observed galaxy clusters often display perturbed morphologies suggesting they underwent recent interactions with less massive subclusters.

A number of galaxy clusters exhibit diffuse radio emission in their periphery, which is not associated with galactic sources \citep[e.g.][]{Feretti1996, vanWeeren2009, Bonafede2012}. These Mpc-scale structures are know as radio relics and they are useful as probes of merger shocks. They are generally interpreted as being driven by shock waves that propagate outwards at supersonic speeds, re-accelerating relativistic electrons through the Fermi mechanism in the very low-density outskirts of the cluster \citep[e.g.][]{Fujita2003, Gabici2003}.

From the theoretical standpoint, idealised numerical simulations of binary cluster mergers supply a wealth of insight into the  outcomes of these events \citep[e.g.][]{Roettiger1993, Schindler1993, Pearce1994, Roettiger1997, Ricker1998, Roettiger2000, Ricker2001, Ritchie2002, Poole2006, ZuHone2009b, ZuHone2010, ZuHone2011}. Hydrodynamical simulations of merging cluster designed to model specific observed objects have frequently focused on the Bullet Cluster \citep[e.g.][]{Takizawa2005, Takizawa2006, Milosavljevic2007, Springel2007, Mastropietro2008}. Recently, \cite{vanWeeren2011} carried out hydrodynamical simulations to model the galaxy cluster \mbox{CIZA~J2242.8+5301} and used its double radio relics to set constraints on the merger geometry, mass ratio and time-scale. \cite{Bruggen2012} modelled \mbox{1RXS~J0603.3+4214}, another radio relic cluster, as a triple merger.

In fully cosmological hydrodynamical simulations, clusters mergers are studied in a more realistic but less well-controlled environment, and sometimes at the cost of lower spatial or mass resolution. They provide useful analyses of the statistical properties of merger shocks, such as cold fronts \citep{Hallman2010} and the distributions of Mach numbers \cite[e.g.][]{Vazza2011, Araya-Melo2012, Planelles2012}. Typical Mach numbers in shocks driven by cluster collisions tend to be $\mathcal{M}\lesssim3$, but strong shocks may arise under some circumstances. For example, \cite{Finoguenov2010} obtained $\mathcal{M}\sim2$ from the density and temperature drops in Abell~3667 and \cite{Markevitch2005} had found a similar value for Abell~520. A shock as strong as $\mathcal{M}\sim4$ was estimated by \cite{vanWeeren2010} from the radio relics of \mbox{CIZA~J2242.8+5301}.

Using Suzaku X-ray observations, \cite{Akamatsu2012b} measured the temperature jump in the western radio relic of Abell~3376, obtaining $\mathcal{M}=2.91\pm0.91$. From polarisation and spectral indices studies of the radio relics of Abell~3376, \mbox{\cite{Kale2011, Kale2012}} estimated  $\mathcal{M}\sim2.2-3.3$. From a cosmological simulation, \cite{Paul2011} were able to identify one merging cluster whose shock structure resembles the radio relics of Abell~3376.

Abell~3376 (hereafter A3376) is a nearby ($z=0.046$) massive galaxy cluster with a distinctive bullet-like morphology, suggesting an ongoing collision taking place. It is possibly the closest cluster exhibiting such a morphology. In its outskirts, a pair of prominent Mpc-scale radio relics are present \citep{Bagchi2006}. Its virial mass is estimated as $5.0\times10^{14}M_{\odot}$ \citep[based on galaxy velocity dispersion]{Girardi1998} and its \mbox{0.1--2.4~keV} luminosity is $L_{X}\simeq2.5\times10^{44}$~erg/s \citep[based on \textit{Rosat} data]{Ebeling1996}. The first and second brightest cluster galaxies (BCG) are separated by $\sim 970 h_{70}^{-1}$~kpc, but it is the second BCG that coincides with the peak of X-ray emission.

\citet{Bagchi2006} have analysed early XMM-\textit{Newton} X-ray and VLA radio data, proposing that the radio relics observed may also be the site of acceleration of very energetic cosmic rays, up to $10^{18}$~eV. This would be possible with first-order Fermi mechanism at the shockwave front. They set forth two possible scenarios that could be responsible for producing the shock: (i) the accretion of intergalactic medium flowing down towards the cluster; or (ii) the collision of two clusters, but the issue remained unresolved.

Our goal in this work is to run $N$-body+SPH simulations tailored for the specific case of A3376 using more recent XMM-\textit{Newton} data in order to constrain the actual dynamical history of this system.

This paper is organized as follows. Simulation techniques and initial conditions are described in Section~\ref{sec:simulations}. X-ray observations are described in Section~\ref{sec:xray}. In Section~\ref{sec:results} we present the global merger evolution, explore the parameter space of different possible collisions, and compare the simulation results to observations; Mach number and dark matter distribution are discussed. Finally, we summarise and conclude in Section~\ref{sec:conclusions}. Throughout this work we assume a standard $\Lambda$CDM cosmology with $\Omega_{\Lambda}=0.7$, $\Omega_{M}=0.3$ and $H_{0}=70$~km~s$^{-1}$~Mpc$^{-1}$.

\section{Simulations}
\label{sec:simulations}

We set up an idealised numerical model to represent the merging of two initially isolated galaxy clusters. The main purpose of these simulations is to reproduce certain features of the galaxy cluster A3376, especially the global morphology of the intracluster medium. Even though this model relies on several simplifications, it allows us to reconstruct a possible scenario for the dynamical history of A3376. By comparing the results of a large set of simulations, it is possible to set approximate constraints on some collision parameters.

\subsection{Techniques}
\label{sec:techniques}

We consider the collision of two spherically symmetric galaxy clusters: a more massive cluster A (the major cluster) and a less massive cluster B (also referred to as the minor cluster or the subcluster). A range of mass ratios $M_{B}/M_{A}$ is explored. Each cluster is composed of dark matter particles and gas particles, and a baryon fraction of 0.18 is adopted throughout.

In all simulations, the major cluster has a total of \mbox{$N_{A}=6\times10^{6}$} particles proportionally divided between dark matter and gas, such that the mass resolution is the same \mbox{$m_{i}=1\times10^{8} M_{\odot}$} for both species. The minor cluster's mass and particle number are scaled down such that all particles have the same mass.

Simulations were performed using the public version of the parallel SPH code Gadget-2 \citep{Springel2005} with a softening length of $\epsilon=5$~kpc and a maximum time step of 0.001~Gyr. The intracluster medium is represented by an ideal gas with adiabatic index $\gamma=5/3$. As a first approximation, cooling is not taken into account in the simulations, since the cooling time-scale is larger than the merger time-scale. Galaxies themselves account for a small fraction of the total cluster mass \cite[$\sim 3\%$, e.g.][]{Lagana2008} and their gravitational influence may be disregarded. Since we do not include stars in these simulations, feedback and star formation are not present. Even though magnetic fields are believed to give rise to the radio relics observed in A3376, they are not expected to play an important role in determining the global cluster morphology, and are ignored in our simulations. The evolution of the system is followed for 5~Gyr, but the relevant phases take place in a time-scale of not more than about 1~Gyr. Because of that, and of the small spatial extent of the system, cosmological expansion is neglected. Simulations were carried out on a 2304-core SGI Altix cluster.

\subsection{Initial conditions and models}
\label{sec:ic}

For the dark matter halo, a \cite{Hernquist1990} density profile is adopted:

\begin{equation}
\rho_{h}(r) = \frac{M_{h}}{2 \pi} ~ \frac{r_{h}}{r~(r+r_{h})^{3}} 
\end{equation}

\noindent where $M_{h}$ is the dark matter halo mass, and $r_{h}$ is a scale length. This is similar to the NFW profile \citep*{NFW1997} except in the outermost parts beyond the $r_{200}$ radius (also understood as the virial radius), with the advantage of having a finite total mass. Conveniently, many of its properties (such as potential, cumulative mass and distribution function) can be expressed analytically.

For the gas distribution, we employ a \cite{Dehnen1993} density profile:

\begin{equation}
\rho_{g}(r) = \frac{(3-\gamma)~M_{g}}{4\pi} ~ \frac{r_{g}}{r^{\gamma}(r+r_{g})^{4-\gamma}} 
\end{equation}

\noindent where $M_{g}$ is the gas mass and $r_{g}$ is a scale length. This has the benefit of preserving the analytical simplicity of several useful quantities (chiefly the derivatives of the potential), while also allowing the possibility of a flat core, which is achieved by setting its $\gamma$ parameter to zero. The resulting profile resembles that of a $\beta$-model \citep{Cavaliere1976} and is thus suitable to represent the intracluster medium of an undisturbed cluster without a pronounced cool-core, i.e. without a steep density profile in the centre. 

To create a numerical realisation of this dark matter + gas system, we set up the initial positions and velocities following the procedure outlined by \cite{Kazantzidis2006}. First, the dark matter cumulative mass function is uniformly sampled in the interval $[0,M_{h}]$ and the inverse function $r(M)$ is used to provide $r$, the distance from the centre, for each dark matter particle. The same is done for the gas particles. From these radii, cartesian coordinates are obtained by assigning random directions to the position vectors.

One possible method to obtain the velocities of the collisionless particles is the so called `local Maxwellian approximation'. It consists in drawing velocities from Maxwellian distributions with dispersions obtained from solving the Jeans equation at each radius \citep{BinneyTremaine}. However, due to the inadequacies of this approach which have been pointed out by \cite{Kazantzidis2004}, one must resort to the distribution function itself. In this way, no assumptions need to be made about the local shape of the velocity distribution (apart from the assumptions of isotropy and spherical symmetry) and the exact distribution function $f(\mathcal{E})$ is given by Eddington's formula \citep{Eddington1916, BinneyTremaine}:

\begin{equation} \label{eq:eddington}
f(\mathcal{E}) =\frac{1}{\sqrt{8} \pi^2}\left[ \int_{0}^{\mathcal{E}} 
\frac{{\rm d}^2 \rho_{h}}{{\rm d} \Psi^2} \frac{{\rm d}
\Psi}{\sqrt{\mathcal{E}-\Psi}} + \frac{1}{\sqrt{\mathcal{E}}} \left (\frac{{\rm d} \rho_{h}}
{{\rm d}\Psi}\right)_{\Psi=0}\right]
\end{equation} 

\noindent where $\Psi=-\Phi = -(\Phi_{h} + \Phi_{g})$ is the relative total potential, and $\mathcal{E}=\Psi-v^{2}/2$ is the relative energy. For physically meaningful $\rho_{h}(r)$ and $\Psi(r)$, the last term in equation~(\ref{eq:eddington}) vanishes. This leaves an integrand that depends on (the second derivative of) the dark matter density expressed as a function of the total potential, $\rho_{h}=\rho_{h}(\Psi)$. The integration is evaluated numerically and the function $f(\mathcal{E})$ is tabulated in a fine grid over a range of energies and then interpolated wherever necessary. Random pairs $(\mathcal{E},f)$ are drawn and values of $v^2$ are accepted according to the \cite{vonNeumann1951} rejection technique. This assigns a speed to each collisionless particle and cartesian velocity components are obtained assuming random directions for the velocity vectors.

Since the gas is assumed to be in hydrostatic equilibrium, each volume element of the fluid is initially at rest. The SPH particles require an additional quantity to be set up: their internal energy (i.e. temperature). From the assumption of hydrostatic equilibrium, it follows that for a chosen gas density $\rho_{g}(r)$, the temperature profile is uniquely specified as:

\begin{equation}
T(r) = \frac{\mu m_{p}}{k} \frac{1}{\rho_{g}(r)} \int_{r}^{\infty} \rho_{g}(r')\frac{GM(r')}{r'^2} \, dr'
\end{equation}

\noindent where $\mu$ is the mean molecular weight, $m_{p}$ is the proton mass, $k$ is the Boltzmann constant and $M(r')=M_{g}(r')+M_{h}(r')$ is the total mass inside the radius $r'$. 

Each cluster model is allowed to relax in isolation for a period of 5~Gyr, typically a few dynamical time-scales, prior to the beginning of the actual collision. This ensures that transient numerical effects, however minor, will have had time to subside. The clusters are then placed at a sufficiently large initial separation $d_{0}$, having an initial relative velocity $v_{0}$ in the direction of the x-axis. To avoid spurious tidal effects in the initial conditions, we employ an initial separation of $d_{0}=4$~Mpc, which is approximately twice as large as the sum of the clusters' virial radii.

\begin{table}
\caption{Inital condition parameters. Model 233 is the fiducial model.}
\label{tb:models}
\begin{center}
\begin{tabular}{c c c c c c r}
\hline
model & label & $\frac{M_{B}}{M_{A}}$  & $\frac{n_{0,B}}{n_{0,A}}$ & $v_{0}$  & $b_{0}$ & $N_{total}$ \\ 
     &       &                &                   & (km/s) & (kpc) &  \\
\hline
231  & mr2   & 1/2 & 4  & 1500 & 0   & $9\times10^{6}$ \\ 
232  & mr4   & 1/4 & 4  & 1500 & 0   & $7.5\times10^{6}$ \\ 
\bf{233}  & \bf{mr6}   & \bf{1/6} & \bf{4}  & \bf{1500} & \bf{0}   & $\mathbf{7\times10^{6}}$ \\ 
234  & mr8   & 1/8 & 4  & 1500 & 0   & $6.75\times10^{6}$ \\ 
~\\
241  & b150  & 1/6 & 4  & 1500 & 150 & $7\times10^{6}$ \\ 
242  & b350  & 1/6 & 4  & 1500 & 350 & $7\times10^{6}$ \\ 
243  & b500  & 1/6 & 4  & 1500 & 500 & $7\times10^{6}$ \\ 
~\\
238  & v500  & 1/6 & 4  & 500  & 0 & $7\times10^{6}$ \\ 
239  & v1000 & 1/6 & 4  & 1000 & 0 & $7\times10^{6}$ \\ 
240  & v2000 & 1/6 & 4  & 2000 & 0 & $7\times10^{6}$ \\ 
~\\
235  & n1    & 1/6 & 1  & 1500  & 0 & $7\times10^{6}$ \\ 
236  & n2    & 1/6 & 2  & 1500 & 0 & $7\times10^{6}$ \\ 
237  & n6    & 1/6 & 6  & 1500 & 0 & $7\times10^{6}$ \\ 
\hline
\end{tabular}
\end{center}
\end{table}

Numerous combinations of plausible initial condition parameters are possible and, in the search for a `best-fitting' model, hundreds of simulations were run. Table \ref{tb:models} lists the initial condition parameters of the sample of models that are reported in this paper. This sample focuses on the variations of four parameters, namely: the total mass ratio $M_{B}/M_{A}$, ranging from 1/2 to 1/8; the ratio of the central gas densities $n_{0,B}/n_{0,A}$ ranging from 1 to 6; the initial relative velocity $v_{0}$, ranging from 500 to 2000~km/s; and the impact parameter $b_{0}$, ranging from 0 to 500~kpc.

The major cluster's total mass is always $M_{A}=6\times10^{14}M_{\odot}$ and its central gas density is $n_{0,A}=1.2\times10^{-2}$cm$^{-3}$ in all cases. These values were chosen to ensure that both the total mass and the total X-ray luminosity of the resulting object are within the same order of magnitude as the observed cluster. The initial relative velocity is always parallel to the $x$-axis, even when a non-zero impact parameter (a shift in the initial position of the subcluster along the $y$-axis direction) is present. 

\section{X-ray data}
\label{sec:xray}

\subsection{Observations}
\label{sec:xrayObs}

The numerical simulations presented here will be compared to archival X-ray observations. A3376 was observed twice by the XMM-\textit{Newton} satellite, in 2003 (revolution 0606, P.I. M.~Markevitch) and in 2007 (revolution 1411, P.I. M.~Johnston-Hollitt). While the 2003 observation was partially analysed by \citet{Bagchi2006} (they did not use the pn detector), the 2007 observation is, as far as we know, unpublished.

Both observations were done in Prime Full Window with ``medium'' filter. We have run the Science Analysis System (SAS\footnote{See http://xmm.esac.esa.int/} 11.0) pipeline, removing bad pixels, electronic noise, and correcting for charge transfer losses. For the EPIC MOS1 and MOS2 cameras we have kept only events with PATTERN $\le$ 12 and FLAG = 0 (events on the field of view). For the pn camera, have kept events with PATTERN $\le$ 4 and FLAG = 0, following the standard procedure recommended by the SAS team. Both observations were screened for high particle background periods. We have constructed light-curves in the [1-12 keV] band and filtered out time intervals of anomalously high flux. The final exposure times for the 2003 observation were 23.0, 22.9, and 16.0~ks for the MOS1, MOS2, and pn, respectively. For the 2007 observation they were 36.8, 39.8, and  25.6~ks for the MOS1, MOS2, and pn, respectively.

When necessary, the background was taken into account by using the publicly available EPIC blank sky templates described by \cite{Read2003}, taking into account the observation mode and filter. Each blank sky background was further normalised using the observed spectrum obtained in an annulus (between 12.5--14.0 arcmin), taking care to avoid the cluster emission. 

\subsection{Analysis}
\label{sec:xrayAnalysis}

With the cleaned event files, we have produced the exposure-map corrected images in the [0.5--8.0 keV] band combining all XMM data available. This image will be compared below to a simulated image from the $N$-body simulation.

We have also produced a 2D temperature map, using an adaptive kernel technique \citep{Durret2008}. A cell of variable size must have a minimum count number (of the order of $10^3$ after background subtraction). A cell that meets this criterium has its spectrum fitted by an absorbed, single temperature MEKAL model \cite[bremsstrahlung and emission lines,][]{Kaastra1993, Liedahl1995} using the XSPEC~12.5 package\footnote{See http://heasarc.gsfc.nasa.gov/docs/xanadu/xspec/} \citep{Arnaud1996}. The free parameters are the intra-cluster plasma temperature and metallicity (metal abundance, mainly iron). 

The spectral fits are done in the [0.7--8.0 keV] band. We fixed the absorption using the Galactic value of the neutral hydrogen column density ($4.6 \times 10^{20}\,$cm$^{-2}$; Leiden/Argentine/Bonn (LAB) Survey), estimated with the \texttt{nh} task from FTOOLS\footnote{See http://heasarc.gsfc.nasa.gov/ftools/}. The effective area files (ARFs) and the response matrices (RMFs) were computed for each cell in the image grid. We produced a temperature map combining both observations, which will be compared to our simulations below.

\section{Results}
\label{sec:results}

\subsection{Global evolution}
\label{sec:evolution}

\begin{figure}
\centering \includegraphics[width=0.85\columnwidth]{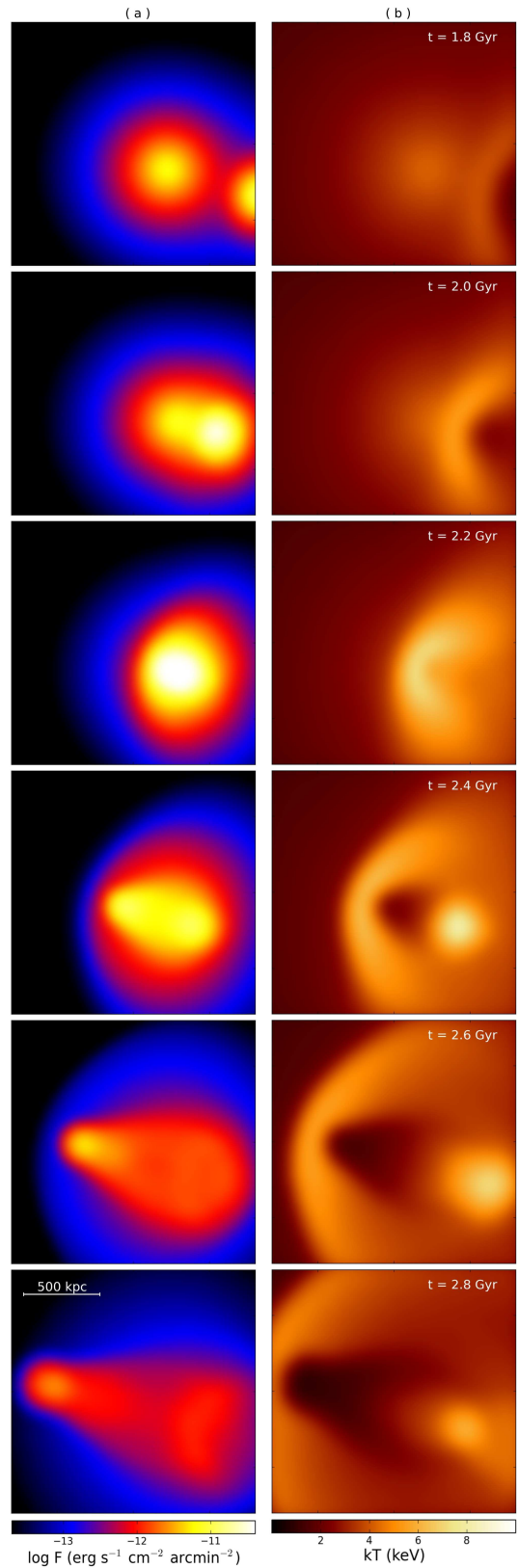}
\caption{Temporal evolution of model 233: (a) projected X-ray surface brightness, in the left column; and (b) projected emission-weighted temperature, in the right column. The six rows displays instants 0.2 Gyr apart. Each frame is 1.6$\times$1.6 Mpc, the spatial scale is the same for all frames, and the surface brightness and temperature ranges are both constant.}
\label{fig:time}
\end{figure}

To illustrate the temporal evolution of a cluster merger simulation, Fig.~\ref{fig:time} displays a sequence of snapshots (of model 233) spaced by 0.2~Gyr, in frames of 1.6$\times$1.6 Mpc. This is a head-on collision along the $x$-axis, which appears arbitrarily rotated on the plane of the sky merely to match the position angle of the observation, a rotation having of course no further relevance. The meaningful orientation is the inclination angle $i$, between the collision axis and the plane of the sky. All frames in Fig.~\ref{fig:time} are projected at a inclination $i=40^{\circ}$, discussed in detail in Section~\ref{sec:inclination}. To allow for a somewhat fair comparison with observations, we compute the projected X-ray surface brightness (left column of Fig.~\ref{fig:time}) and the projected emission-weighted temperature (right column of Fig.~\ref{fig:time}).

Even though the simulations themselves are adiabatic, when generating these simulated images we assume the X-ray radiative losses can be described by a cooling function that takes more than just bremsstrahlung into account. From the simulation output, we estimate the emission using a cooling function $\Lambda(T)$ that was computed using the MEKAL emission spectrum model with the XSPEC 12.5 package, appropriate for a plasma of
metallicity equal to 0.3~$Z_{\odot}$. At high energies the free-free emission $(\propto n_{e}^{2}\sqrt{T})$ is predominant (and that is the quantity usually employed as a proxy for emission in galaxy cluster simulations), but at low energies collisional excitation dominates. The emission is then projected along the line of sight to give the X-ray surface brightness maps. Temperature maps are generated by weighting particle temperatures by their emission, and projecting them along the line of sight. 

In Fig.~\ref{fig:time} the subcluster comes from the lower right corner of the frame towards the major cluster. Most of the relevant dynamical evolution of the merger takes place within an interval of 1~Gyr. The simulation starts at $t=0$ when the clusters are 4~Mpc apart, and central passage occurs at $t\simeq2.15$~Gyr. Approximately 0.5~Gyr after central passage a configuration is reached in which the overall gas morphology approximates fairly well the shape of the observed cluster. At this moment, the subcluster has passed beyond the major cluster centre while remaining significantly denser, which makes the emission slightly more intense in the region of the nose of the shock than in the gas behind it. At earlier times no such distinction is noticeable, whereas at later times the global shape is excessively elongated and a detachment develops between the subcluster core and the gas left behind.

In this particular model, the subcluster has 1/6 of the main cluster's total mass, but the gas in its core is more centrally concentrated, being denser by a factor of 4. The subcluster is also colder, having an initial central temperature of $\sim 1$~keV, while the major cluster has $4-5$~keV. The heated gas ahead of the cold subcluster is visible as a bow shock in the right column of Fig.~\ref{fig:time} and it is further discussed in Section~\ref{sec:mach}.

At the instant of best match ($t=2.625$~Gyr for model 233) the system evidently lacks spherical symmetry. Nevertheless we measure the spherically averaged density profile, centred on the point of highest total density. This is located at the major cluster's dark matter peak, around which the bulk of the total mass still lies (see Section~\ref{sec:darkmatter}). We obtain the radius $r_{200}\simeq1.5$~Mpc, at which the mean density has dropped to 200 times the critical density $\rho_{c}$. This radius also encompasses the minor cluster's dark matter peak, as well as all the gaseous structures discussed here. The mass enclosed within $r_{200}$ is $M_{200}\simeq4.0\times10^{14}M_{\odot}$. The integrated X-ray luminosity (within $r_{200}$ and in the energy range 0.1--2.4~keV) is $L_{X}\simeq4.3\times10^{44}$~erg/s if the emission comes solely from bremsstrahlung, or $L_{X}\simeq6.8\times10^{44}$~erg/s if metals and line emission are also taken into account.

\subsection{Exploration of parameter space}
\label{sec:parameterspace}

In order to attempt to constrain some of the merger parameters, we explored numerous different initial condition configurations, in search of a `best-fitting' model. Such a model would have to simultaneously satisfy, to an approximate degree, the following criteria: the overall gas morphology should be reminiscent of the X-ray observation of A3376; the temperature should be in the appropriate observed range; the total mass and total luminosity should fall in the same order of magnitude as those inferred from observations. A possibly important additional criterium regarding the distance between the two brightest cluster galaxies is discussed in Section~\ref{sec:darkmatter}. We take the BCGs separation to be a proxy for the separation between the two dark matter peaks. While model 233 may not necessarily be the one that strictly optimises each single criterium, it is the one that provided the most acceptable compromise among them. It is of course not possible to rule out the existence of alternative combinations of parameters that might result in similarly acceptable models. We explored physically motivated ranges of parameters (albeit in a limited number of combinations due to the computational cost) and, at least for these ranges, certain configurations may me reliably excluded.

As far as morphological comparison is concerned, we refrain from attempts to implement quantitative algorithms. Instead we rely on visual inspection, which may not be the most objective approach and conclusions drawn from it ought to be regarded carefully. Nevertheless, this is an approach not without its merits. As exemplified by the age-tested practice of morphological classification of galaxies, visual inspection tends to yield surprisingly reliable results, which are often difficult to reproduce algorithmically. The type of comparisons we carry out here fall within this effort of making approximate judgements of morphology by eye, taking into account both the overall appearance and various detailed aspects. While pixel-by-pixel subtraction (or some variation of it) could provide quantitative figures, it could hardly be expected to be sensitive to the same number of visual cues and at the same level of flexibility that evaluation by eye can afford.

\begin{figure*}
\includegraphics[width=\textwidth]{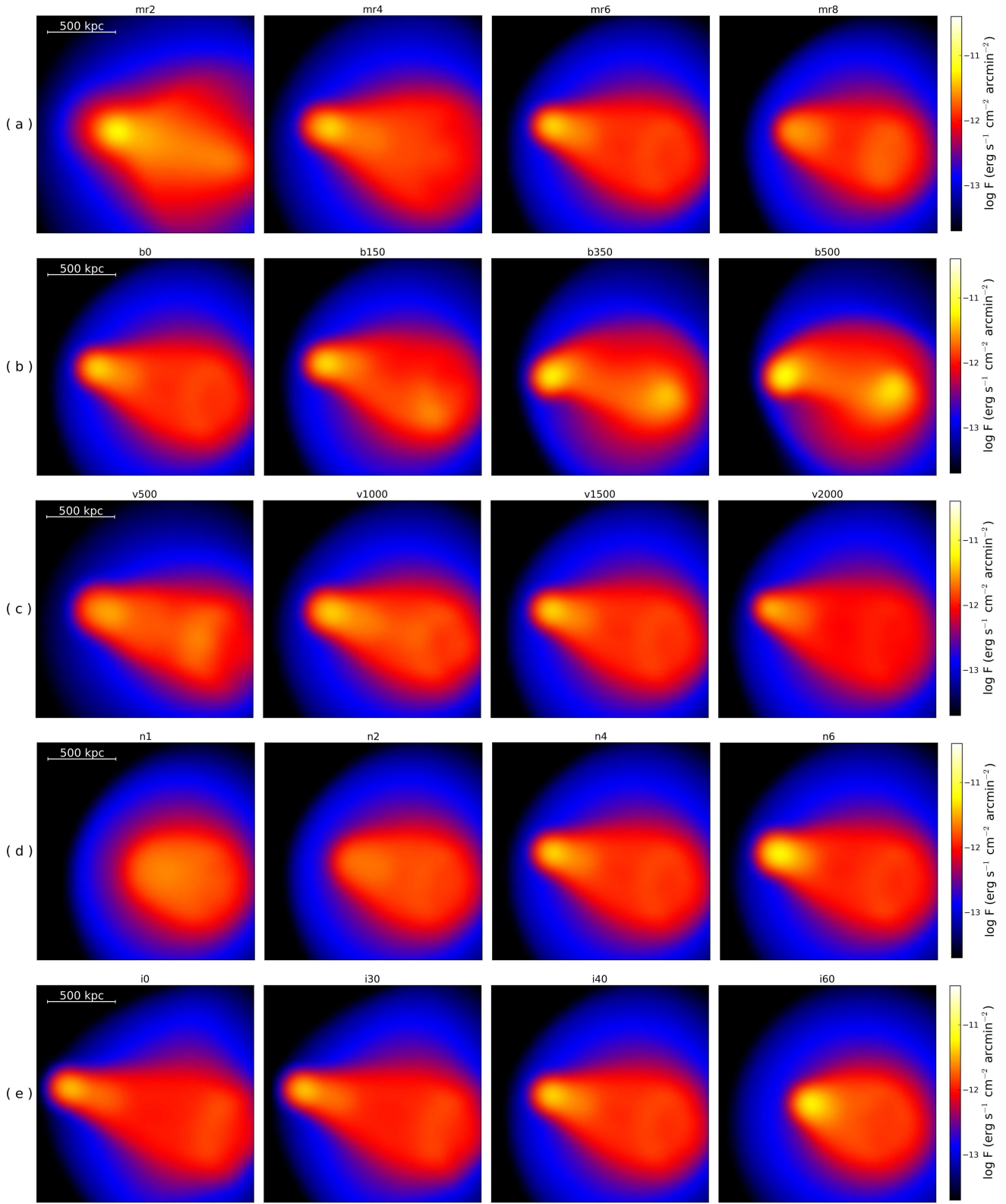}
\caption{Projected X-ray surface brightness for different models. Each row displays four variations of one given parameter: (a) mass ratio; (b) impact parameter; (c) initial relative velocity; (d) relative gas concentrations; (e) inclination.}
\label{fig:variations}
\end{figure*}

Here we present a systematic comparison of a sample of the models we explored. This is meant to allow a comparison of the X-ray morphology of the models listed in Table~\ref{tb:models}. Taking model 233 to be the standard, Fig.~\ref{fig:variations} displays variations around that model. The surface brightness scale and spatial scale are the same as in Fig.~\ref{fig:time}. Each row in Fig.~\ref{fig:variations} displays variations of one given parameter: (a) mass ratio; (b) impact parameter; (c) initial relative velocity; (d) relative central gas densities; and (e) inclination. For each of these properties, four variants are given, one of them being the fiducial model itself. This highlights the individual effect of varying each parameter separately. With the exception of row (e), every model is shown with the same inclination $i=40^{\circ}$ to allow for a fair comparison. All snapshots are shown at the same instant $t=2.625$~Gyr, with the exception of row (c), because models with different initial velocities have substantially different time-scales. 

\subsubsection{Mass ratio}

Row (a) of Fig.~\ref{fig:variations} shows the outcome of cluster mergers having mass ratios of 1/2, 1/4, 1/6 and 1/8. It is clear that major mergers result in a quite distinct morphology. Models having $M_{B}/M_{A}$ in the range of 1/6 -- 1/8 are to be preferred. \cite{Neistein2008}, based on the extended Press-Schechter formalism, provide estimates of the rate of mergers having mass ratio above a certain value at a given redshift. A cluster of model 233's $M_{200}$ at A3376's redshift will have undergone one merger per $\sim4.8$~Gyr having \mbox{$M_{B}/M_{A}>1/6$}.

\subsubsection{Impact parameter}

Typical impact parameters in galaxy cluster mergers are of the order of a few 100~kpc \citep{Sarazin2002, Ricker1998,  Ricker2001}. The four models shown in row (b) of Fig.~\ref{fig:variations} have initial impact parameters $b_{0} = 0, 150, 350$~and~$500$~kpc. The most evident effect of off-centre mergers is the loss of symmetry around the collision axis. The curved trajectory of the subcluster is partly responsible for the distinctive shape of the resulting objects. In a head-on collision, the major cluster's central gas is spread out as the denser subcluster passes through it, decreasing its density. If however the two clusters don't pass through each other's centre, the major cluster's core remains relatively undisturbed, and as a result two separate clumps of dense gas are still visible. The asymmetry due to a curved trajectory could be hidden from view if the orbital plane were seen exactly edge-on. But even under that particular configuration, both cores would still be discernible. Since A3376 displays neither noticeable asymmetry nor two cores, there is no reason to go beyond the scenario of a head-on collision.

It should be noted that $b_{0}$ refers to a shift in the direction of $y$-axis in the initial conditions, i.e. when the clusters are 4~Mpc apart in the $x$-axis direction. The minimum separation $b_{min}$, which is the distance between the cluster centres at the instant of closest approach, is considerably smaller. For the three off-axis models presented in Fig.~\ref{fig:variations}b, the approximate distances at pericentric passage are respectively $b_{min}=100,200$~and~$250$~kpc. This sets a tight constraint, since a $b_{min}$ as small as 100~kpc would be sufficient to produce noticeable asymmetry.

\subsubsection{Initial relative velocity}

Given the major cluster mass $M_{A}$, the subcluster's free fall velocity at $d_{0}=4$~Mpc would be approximately $v_{0}=1250$~km/s, if it were a point mass having been released from infinity at rest.

The models displayed in row (c) of Fig.~\ref{fig:variations} have initial relative velocities $v_{0}=500, 1000, 1500$~and~$2000$~km/s. Because the time-scales are not the same, they are compared at different instants, each chosen to be that which best resembles A3376. As far as the gas morphology is concerned, the different velocities affect the sharpness of the nose of the shock. Models v1000 and v2000 were dismissed not on account of their X-ray morphologies, which are not inadequate. However, model v2000 gives rise to excessively high temperatures in the shock region. In model v1000, on the other hand, it takes a long time for the dark matter peaks to be sufficiently separated (see Section~\ref{sec:darkmatter}) and by that time the morphology has deteriorated. The 1500~km/s model provides a tolerable compromise between morphology, temperature and dark matter peaks separation. The Mach numbers corresponding to these velocities are discussed in Section~\ref{sec:mach}. 

\begin{figure}
\includegraphics[width=\columnwidth]{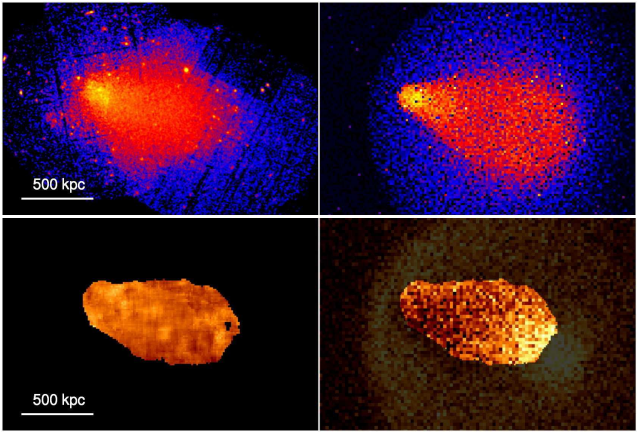}
\caption{Comparison between observations of A3376 (left) and model 233 (right). The upper row displays the XMM-\textit{Newton} X-ray surface brightness in the [0.5-8.0 keV] band compared to a simulated image whose resolution has been deliberately degraded. In the lower row, the left panel shows the observed temperature maps. In the simulated temperature map, a semi transparent mask highlights the region in which there is data.}
\label{fig:comparison}
\end{figure}

\begin{figure}
\includegraphics[width=\columnwidth]{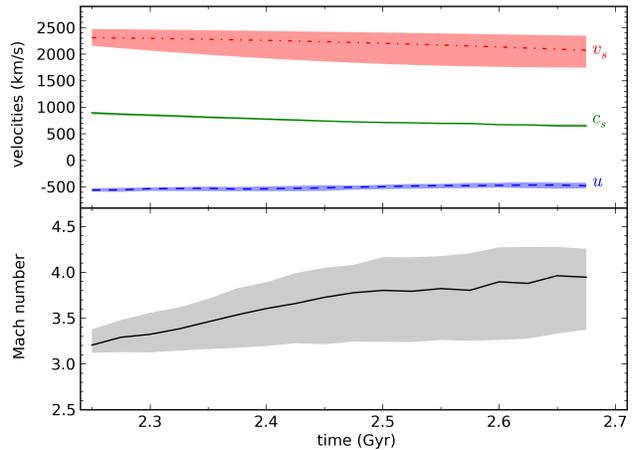}
\caption{The upper panel displays the shock velocity $v_{s}$ (dot-dashed line) and the velocity of the upstream gas $u$ (dashed line), both measured in the centre-of-mass rest frame; also shown is the sound speed $c_{s}$ (solid line). The lower panel shows the resulting Mach number as a function of time for this model (233). The shaded regions represent upper and lower boundaries for each quantity, computed from models v1000 and v2000 (and shifted to match the same time interval).}
\label{fig:mach2}
\end{figure}

\subsubsection{Central gas density}

Row (d) of Fig.~\ref{fig:variations} highlights the effects of relative central gas densities, i.e. the effects of the concentration of the subcluster. It shows models with $n_{0,B}/n_{0,A}=1, 2, 4$~and~$6$. There is a clear dependence on the concentration of the subcluster. When the clusters have comparable concentrations, the subcluster gas is hardly distinguishable from the major cluster gas. The outcome is a somewhat uniform emission with no X-ray peak. If the subcluster is considerably denser, it is able to cross the major cluster core and to remain relatively cohesive as it emerges. Because in model n6 the X-ray peak is excessively prominent,  the preferred model is that in which the subcluster central gas is 4 times denser the major cluster's.

\subsubsection{Inclination}
\label{sec:inclination}

Row (e) of Fig.~\ref{fig:variations} shows model 233 in the same instant projected under four different inclinations $i=0^{\circ}, 30^{\circ}, 40^{\circ}$~and~$60^{\circ}$. For low inclinations, the shape is excessively elongated at this time. For very high inclinations it is excessively round. Unfortunately the observations provide no information on the spatial orientation of the cluster. Here again, the best configuration is chosen not solely from the X-ray morphology, although it is a good indicator. For example, at later time an inclination of $i=60^{\circ}$ provides a morphology that is also acceptable. However the projected separation of the dark matter peaks is very time dependent and also very sensitive to inclination (see Section~\ref{sec:darkmatter}). By the time the dark matter peaks are sufficiently separated with $i=60^{\circ}$, the morphology is no longer adequate. Therefore the $i=40^{\circ}$ projection provides the best compromise.

\subsection{Comparison to observations}
\label{sec:observations}

A comparison between observations of A3376 and model 233 is given in Fig.~\ref{fig:comparison}. Despite the high-resolution of the numerical model itself, the simulated images are deliberately degraded by undersampling the particles, by applying a gaussian smoothing and by the addition of noise. 

In the observed temperature map (see Section~\ref{sec:xrayAnalysis}, there is data only within a relatively small region approximately 20~arcmin wide. To make the comparison more straightforward, a region of the same shape is overlaid on the simulated temperature map, while the surrounding area is made semi opaque. This emphasises the region in which data exists, and underscores the simulated temperature features which are not observationally available. Apart from the range of values, the observed temperature map does not exhibit remarkable features that could impose particularly strong constraints on the simulations parameters. We were able to rule out collisions that heated the gas to above 10~keV, for example, and this was used to constrain velocities and concentrations. Nevertheless, the small scale details of the observed temperature map are not reproduced, probably due to the simplifying assumptions of our adiabatic numerical models, which take into account neither substructures nor the various galaxy-related physical processes that might affect the intracluster gas, such as feedback from supernovae and AGN. 

\subsection{Mach number}
\label{sec:mach}

In a cluster merger simulation, the Mach number can be directly measured, as all velocity information is available. One of the most pronounced features is the temperature drop after the bow shock. The successive positions of this discontinuity are used to directly compute the shock velocity $v_{s}$ (in the centre-of-mass rest frame). As shown by \cite{Springel2007} in simulations of 1E0657-56, the gas ahead of the shock is not at rest. In such mergers, the upstream gas is in fact falling towards the incoming subcluster with a considerable velocity $u$ (in the centre-of-mass rest frame). Therefore, the effective relative velocity with which the shock front encounters the pre-shock gas is $v_{s}-u$. Consequently, the Mach number ought to be computed as

\begin{equation}
\mathcal{M} =  \frac{v_{s}-u}{c_{s}}
\end{equation}

\noindent where $c_{s}^{2}=\frac{\gamma k T}{\mu m_{p}}$ is the sound speed of the pre-shock gas. Both the upstream velocity and the sound speed are measured in the region immediately ahead of the shock.

\begin{figure}
\centering \includegraphics[width=0.985\columnwidth]{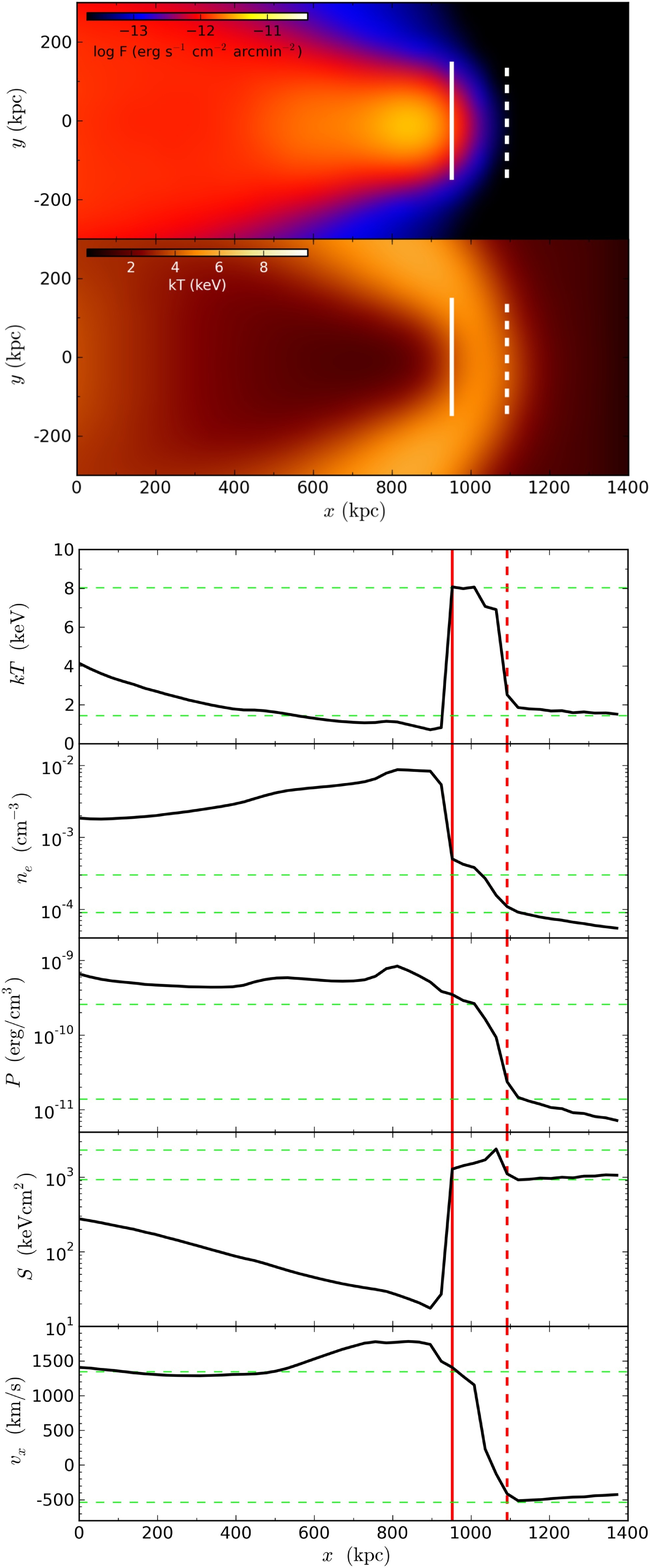}
\caption{Properties of the shock for model 233 at $t=3.3$~Gyr. The first and second panels show respectively the X-ray surface brightness map and the emission-weighted temperature map. The third to seventh panels display the following quantities measured along the $x$-axis: temperature, electron number density, pressure, entropy, and gas streaming velocity. The location of the contact discontinuity is marked by solid vertical lines; the location of the bow shock is marked by dashed vertical lines. The dashed horizontal lines give the values of each quantity before and after the shock, where the expected drops were computed from the Rankine-Hugoniot conditions using the measured Mach number.}
\label{fig:drops}
\end{figure}

For model 233 at $t=2.625$~Gyr the shock front moves forward with $v_{s}=2114$~km/s while the upstream gas falls back at $u=-468$~km/s. The sound speed $c_{s}=666$~km/s implies $\mathcal{M}=3.9$ at this instant. Figure~\ref{fig:mach2} shows how these quantities evolve over an interval of 0.4~Gyr. To provide an indication of the typical ranges of these velocities, the shaded areas represent the boundaries given by models v1000 and v2000 (shifted to match their respective time-scales to that of model 233). The resulting Mach numbers, bounded by these extreme cases, would be in the range of roughly $\mathcal{M}=3-4$.

Once $\mathcal{M}$ has been computed directly from the velocities, it may be used to obtain the expected shock discontinuities from the Rankine-Hugoniot conditions. Figure~\ref{fig:drops} displays five quantities measured along the collision axis: temperature; electron number density; pressure; entropy; and gas streaming velocity in the $x$-axis direction. As a proxy for entropy, the conventional definition $S = k~T~n_{e}^{-2/3}$ is adopted. All of these profiles have been measured using not the projected images, but using the particles within a cylinder of radius 150~kpc passing through the nose of the shock. The upper panels of Fig.~\ref{fig:drops} display the surface brightness and the temperature both projected under $i=0^{\circ}$ inclination. The vertical dashed lines at $x=1148$~kpc mark the position of the shock front, which has been determined as the point where the temperature drop is most intense. The vertical solid lines at $x=980$~kpc mark the position of the contact discontinuity (the cold front), the point where the density drops the most. At the contact discontinuity velocity and thermal pressure are continuous. The horizontal lines indicate the expected drops in each of the five quantities at the shock position, as computed from the Rankine-Hugoniot conditions using the measured $\mathcal{M}$. Assuming $\gamma=5/3$ throughout, the relations between the pre-shock (subscript 1) and post-shock (subscript 2) quantities are \citep[e.g.][]{Landau1959, Shu1992}:

\begin{eqnarray}
\frac{T_{2}}{T_{1}} &=& \frac{5\mathcal{M}^{4}+14\mathcal{M}^{2}-3}{16\mathcal{M}^{2}}\\
\frac{n_{2}}{n_{1}} &=& \frac{4\mathcal{M}^{2}}{\mathcal{M}^{2}+3}\\
\frac{P_{2}}{P_{1}} &=& \frac{5\mathcal{M}^{2}-1}{4}\\
\frac{S_{2}}{S_{1}} &=& \left(\frac{5\mathcal{M}^{2}-1}{4}\right) \left(\frac{4\mathcal{M}^{2}}{\mathcal{M}^{2}+3}\right)^{-5/3}\\
\left( v_{x,2}-v_{x,1} \right) &=& \left( v_{s}-u \right)~\left( \frac{3\mathcal{M}^{2}-3}{4\mathcal{M}^{2}} \right)
\end{eqnarray}

\begin{figure}
\includegraphics[width=\columnwidth]{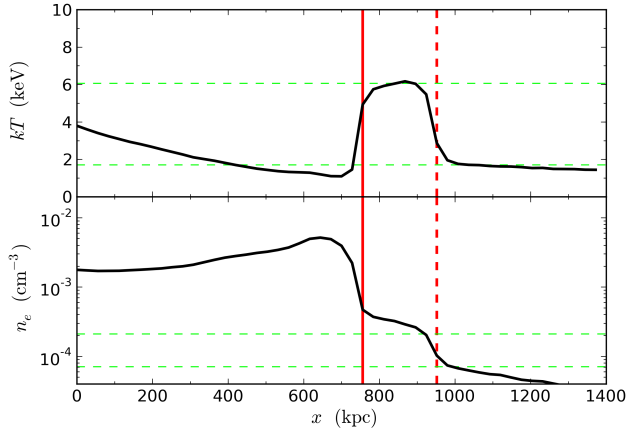}
\caption{Temperature and density measured along the $x$-axis for model 233 seen in projection with inclination $i=40^{\circ}$. If the temperature drop is measured from the inclined model, the inferred Mach number is smaller than the actual value.}
\label{fig:dropsi}
\end{figure}

The good match between the expected drops and the measured profiles indicates that the Mach number could, in principle, be inferred from these quantities. For an observed shock, there is however the difficulty introduced by the unknown inclination. To illustrate this problem, we take the same model 233 at the same instant in time and now project it under the inclination $i=40^{\circ}$. If we now try to measure the temperature and density along the projected collision axis, we obtain the profiles shown in Fig.~\ref{fig:dropsi}. The result is that the height of the temperature peak is lowered because, under projection, the thin region of very hot gas is seen as spread over a larger area. Furthermore, the sharpness of the discontinuities is attenuated. Now, instead of using the known Mach number to compute the drops, we conversely measure the temperature and density drops (horizontal lines in Fig.~\ref{fig:dropsi}) at the shock position. These ratios would lead to $\mathcal{M}=2.9$ for this shock. A Mach number inferred in this manner for a cluster of unknown inclination should thus be regarded as a lower limit. 

\begin{figure}
\begin{flushleft}
\includegraphics[width=0.9\columnwidth]{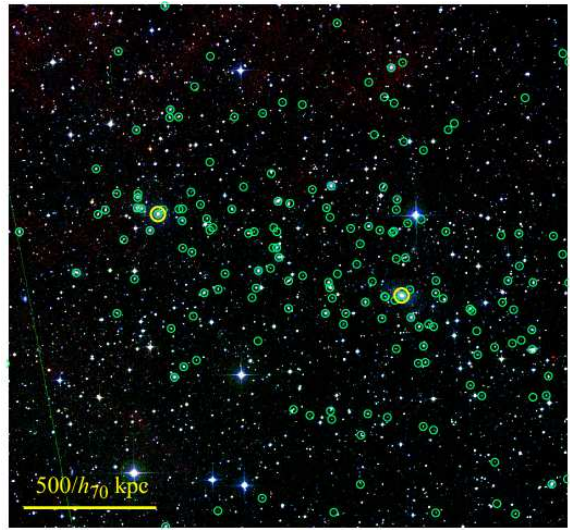}\\
\includegraphics[width=\columnwidth]{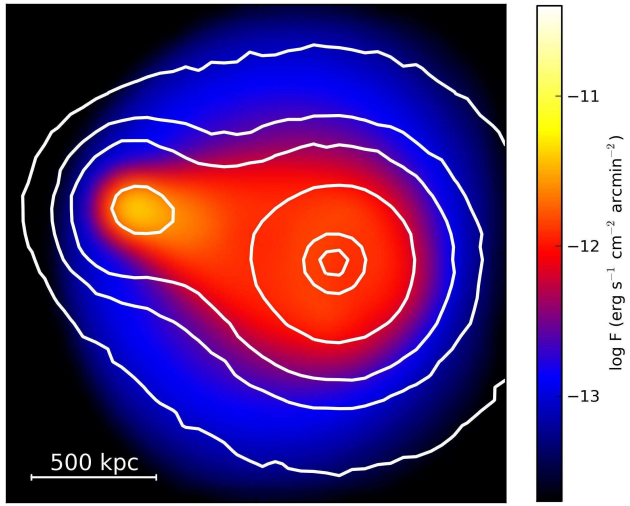}
\end{flushleft}
\caption{The upper panel shows the DSS optical image in which the circles mark positions of the galaxies at the cluster redshift. The first BCG (right) and the second BCG (left) are highlighted. In the lower panel, white countour lines show the total projected mass, overlaid on the projected X-ray surface brightness, for model 233 at $t=2.625$~Gyr with $i=40^{\circ}$.}
\label{fig:contours}
\end{figure}

\subsection{Dark matter distribution}
\label{sec:darkmatter}

In A3376 the brightest cluster galaxy (first BCG) is located not at the peak of X-ray emission, but at a distance of approximately $970 h_{70}^{-1}$~kpc from it. The X-ray peak coincides with the second BCG. A possible interpretation of this peculiarity is a scenario in which a less massive cluster comes from the southwest hosting the second BCG and overtakes the major cluster's core, where the first BCG lies. Our best model accounts for this feature, in the following sense. As there are no galaxies in our simulations, we assume the BCG separation may be identified with the dark matter peaks separation. In the absence of additional clues, it is reasonable to assume that the most massive galaxies should in principle be located at the bottoms of the potential well, i.e. at or near the centroids of the two dark matter peaks.

The lower panel of Fig.~\ref{fig:contours} shows white contour lines that represent the total projected mass for model 233, overlaid on the X-ray surface brightness. The separation between the dark matter peaks is approximately 800--850~kpc at the best-fitting instant. The upper panel of Fig.~\ref{fig:contours} shows an optical `true-colour' (IR-Red-Blue) DSS image of A3376 in which the galaxies at the cluster redshift (selected using NED\footnote{NASA/IPAC Extragalactic Database, http://ned.ipac.caltech.edu/}) are marked by circles and the two BCGs are highlighted. The dark matter separation of model 233 matches the observed BCG distance to within $\sim12\%$. Again, an even better match would be achieved at a slightly later time, but at the price of a deteriorating gas morphology. 

Of course, a possible mass map of A3376 exhibiting two major mass concentrations around the locations of the BCGs would lend more credence to this scenario. Mapping the dark matter distribution by means of gravitational weak lensing is particularly challenging in the case of A3376 due to its proximity. Once or if such map is available, it might either corroborate this scenario or overthrow it. If the dark matter separation turns out to be similar to the BCG separation, the mass map might set more stringent constraints on the simulation parameters. If, on the other hand, a more complicated dark matter structure is revealed, then alternative models will have to be sought out.

An additional degree of freedom that has not been explored in depth in this paper is the relative concentration of the dark matter haloes. For example, if the dark matter is too centrally concentrated, gas temperatures need to be in excess of 10~keV even in the initial conditions to satisfy hydrostatic equilibrium. To obtain physically plausible temperatures of 5~keV in the major cluster, we employ halo scale lengths $r_{h}\sim500$~kpc comparable to the gas density scale lengths $r_{g}$. If the dark matter concentrations are too low, the subcluster's dark matter is able to advance further than its gas and a dissociation develops. As of yet, there is no reason to believe that this is the case for A3376. If a dark matter/gas offset is shown to exist, then it would have to be accommodated by exploring different dark matter concentrations. A more systematic analysis of the dark matter distribution and its dependence on merger parameters is beyond the scope of this paper.

\section{Summary and conclusions}
\label{sec:conclusions}

The peculiar cometary shape of A3376 is suggestive that it has undergone a recent merging event. This is, to best of our knowledge, the closest galaxy cluster displaying such morphology. Furthermore, the diffuse radio emission in the form of double Mpc-scale arcs in its periphery is understood to be caused by shock-accelerated relativistic electrons \citep{Bagchi2006}. 

Using cosmological simulations, \cite{Paul2011} obtains a cluster whose shock wave structure resembles the radio relics of A3376. Dedicated hydrodynamical simulations meant to model one specific merging cluster have been mostly focused on the Bullet Cluster itself \citep[e.g.][]{Milosavljevic2007, Springel2007, Mastropietro2008}. Recently \cite{vanWeeren2011} and \cite{Bruggen2012} carried out simulations of the galaxy clusters \mbox{CIZA~J2242.8+5301} and \mbox{1RXS~J0603.3+4214}, respectively, both of which also exhibit radio relics. In the case of A3376, the morphology is sufficiently uncomplicated that it may be satisfactorily modelled as the encounter of only two objects, thus rendering the reconstruction of its dynamical history relatively simpler.

The simple numerical model we set up in order to investigate the problem consists in the collision of two spherically symmetric galaxy clusters. Equilibrium initial conditions are prepared in which the galaxy clusters are represented by a gas component and a dark matter component. The hydrodynamical simulations themselves are adiabatic, and we assume that radiative losses are unimportant during the time span of the simulations. Yet we use the output of the simulations to generate maps of projected X-ray surface brightness and maps of emission-weighted temperature, in order to compare them to XMM data. 

Starting from a separation of 4~Mpc and a relative initial velocity of 1500~km/s, the clusters meet on a head-on collision and the best-matching moment is reached approximately 0.5~Gyr after central passage. In order to constrain some of the collision parameters, we covered as much parameter space as allowed by the computationally intensive nature of such effort. Ideally, in order to reach a `best model' a number of criteria would have to be simultaneously met, namely: overall gas morphology, temperature, virial mass, total X-ray luminosity, and distance between the two dark matter peaks. While the best model presented here may not necessarily optimise each of these criteria individually, it provided the most adequate compromise. It is of course impossible to argue for the uniqueness of a solution, as alternative combinations of parameters could conceivably provide similar outcomes. This impossibility notwithstanding, the physically motivated ranges of values explored here at least allow us to reliably rule out certain combinations of parameters.

Here we summarise the five main parameters that have been constrained and give approximate estimates of the best ranges: (a) We find that the best matches are obtained for mass ratios in the interval 1/6 -- 1/8. Major mergers are excluded due to their global morphology, while in minor mergers of very small mass ratio, the X-ray peak is not sufficiently intense. (b) Large impact parameters are straightforwardly ruled out as they lead to obvious asymmetries that are not observed in A3376. An approximate upper limit to $b_{0}$ is set at 150~kpc, which implies a separation of $<100$~kpc at pericentric passage. (c) The collision velocity is constrained not by the morphology alone, but also by the temperatures and dark matter peak separation. For initial velocities of 2000~km/s the resulting temperatures are excessively high in the shock region, whereas for 1000~km/s the desired dark matter peak separation takes too long to be reached. (d) An important parameter is the relative central gas concentration. We find the most adequate morphology is obtained when the subcluster is denser than the major cluster by a factor of about 4 in the centre. This determines essentially the prominence of the X-ray peak, which is excessively outstanding if the subcluster central density is too high, or nearly unnoticeable otherwise. (e) Finally, the inclination (the angle between the collision axis and the plane of the sky) plays an important role in the morphology and it is strongly time-dependent. Both the temperature and the total X-ray luminosity are greatly increased at central passage. By the time they reach adequate levels, the shape of the gas distribution is excessively elongated when viewed on the collision plane. We find that, projected under an inclination angle of $i=40^{\circ}$, the morphology is considerably less elongated and the separation between the two dark matter peaks is comparable to the BCG separation to within $\sim12\%$.

Merging clusters generally exhibit complicated temperature structures. The observed temperature map, in a region approximately 20~arcmin wide, shows a mean temperature of $\sim3.5$~keV, but no discernible features that could be used to set strong constraints on the simulations. We use it to rule out models in which the shock heats the gas considerably above the observed range. The outcome of the simulations suggests that the region in which there is data encompasses the contact discontinuity at most, but excludes the shock front itself, i.e. the shock-heated gas ahead of the subcluster's X-ray peak. Furthermore, the small scale details of the observed temperature map are not reproduced in our simulations. If they are the result of interactions between galaxies and the intracluster medium, they could not have been recovered by our simplified simulations which include no such physical processes. The simplifying assumptions of these simulations also exclude the effects of additional substructure.

Cluster mergers are expected to drive supersonic shock waves of typical Mach numbers $\mathcal{M}\lesssim3$ \citep{Sarazin2002} but stronger shocks may arise under some circumstances \cite[e.g.][]{Vazza2011,Planelles2012}. From Suzaku X-ray observations of A3376, \cite{Akamatsu2012b} were able to measure a temperature jump in the western radio relic leading to $\mathcal{M}=2.91\pm0.91$. In our simulations, a supersonic shock wave develops ahead of the colder subcluster. The peak of X-ray emission corresponds to a low temperature, dense gas ahead of which the shock front is located. The layer of heated gas between the edge of the subcluster and the bow shock is $\sim170$~kpc thick. We determine the shock velocity as in \cite{Springel2007}, taking into account the velocity of the upstream gas that is falling towards the incoming subcluster. The actual shock velocity, i.e. the velocity with which the shock wave meets the pre-shock gas, is $\sim2600$~km/s, resulting in a mach number of 3.9. This was computed from the intrinsic properties of the simulation output. If, alternatively, we use the projected images (with inclination $i=40^{\circ}$) to measure the apparent jumps in the temperature and density profiles, we obtain $\mathcal{M}=2.9$ which should then be regarded as a lower limit.

Besides the notorious bullet cluster \citep{Clowe2006}, there is a growing list of merging clusters that have been shown to have an offset between their dark matter and their gas \citep[e.g.][]{Mahdavi2007, Bradac2008, Okabe2011, Dawson2012, Ragozzine2012}. In our best fitting simulations, no substantial dissociation between gas and dark matter developed, and no such models were pursued because, as of yet, there is no observational evidence that this type of offset took place in A3376. Since the total mass distribution of A3376 is unknown, we use the locations of the two brightest cluster galaxies as indicators of the positions of the dark matter peaks. In the absence of further data, it is reasonable to assume that the BCGs should in principle coincide with the centroids of the two dark matter haloes. In this scenario, the major cluster is assumed to host the first BCG, and the subcluster is believed to carry the second BCG, whose position coincides with the peak of X-ray emission. If this is so, the bulk of the dark matter in the system should be located $\sim970 h_{70}^{-1}$kpc away from the X-ray peak. This is what we find in our simulations: a secondary dark matter peak at the position of the X-ray peak, and the main dark matter peak $\sim850$~kpc behind it. The dynamics of how the two dark matter haloes go through each other depends on their relative concentrations and has relevant effects on the final locations of the dark matter peaks. However, a systematic analysis of this aspect will be presented elsewhere.

Determining the projected mass of A3376 by the technique of gravitational weak lensing is difficult because the cluster distance is relatively small. If this proves to be feasible, the resulting mass map could conceivably set tighter or additional constraints on the current merger parameters. If an unexpected dark matter distribution is uncovered, then the current scenario might turn out to be insufficient and a further exploration of the least robust parameters will be required. 

Bearing in mind the limitations of this approach by $N$-body simulations, we have proposed a specific scenario of a dynamical history for the merging event of A3376 and offered a possible combination of parameters that accounts for several of its features. Future weak lensing analysis might either help corroborate this picture or necessitate its improvement.

\section*{Acknowledgements}

This work has made use of the computing facilities of the Laboratory of Astroinformatics (IAG/USP, NAT/Unicsul), whose purchase was made possible by the Brazilian agency FAPESP (grant 2009/54006-4) and the INCT-A. Simulations were also carried out at the Laborat\'orio de Computa\c c\~ao Cient\'ifica Avan\c cada (USP). The authors acknowledge support from FAPESP (2010/12277-9), from the CAPES/COFECUB cooperation and from CNPq. 

\bibliographystyle{mn2e.bst}
\bibliography{myreferences.bib}

\bsp

\label{lastpage}

\end{document}